\documentclass[mathleft
]{an}
\usepackage{graphicx}
\usepackage{times}
\usepackage{listings}
\usepackage[font={small},labelfont=bf]{caption}[2003/12/20]
\usepackage[totalwidth=17cm,totalheight=24.1cm,centering,dvips,pdftex]{geometry}

\begin{document}

\Pagespan{0}{}
\Yearpublication{2009}%
\Yearsubmission{2008}%
\Month{}%
\Volume{}%
\Issue{}%

\title{A standard transformation from XML to RDF via XSLT}

\author{F. Breitling\thanks{\email{fbreitling@aip.de}\newline}
}
\titlerunning{A standard transformation from XML to RDF via XSLT}
\authorrunning{F. Breitling}
\institute{
Astrophysikalisches Institut Potsdam, An der Sternwarte 16, 
D-14482 Potsdam, Germany}

\received{Nov 2008}
\accepted{Jun 2009}
\publonline{later}


\abstract{A generic transformation of XML data into the Resource Description
          Framework (RDF) and its implementation by XSLT transformations is
          presented. It was developed by the grid integration project for
          robotic telescopes of AstroGrid-D to provide network communication
          through the Remote Telescope Markup Language (RTML) to its RDF based
          information service. The transformation's generality is explained by
          this example. It automates the transformation of XML data into RDF
          and thus solves this problem of semantic computing. Its design also
          permits the inverse transformation but this is not yet implemented.}

\maketitle

\tolerance=10000

\section{Introduction}
Progress in astronomy has required experiments of ever increasing complexity.
Modern experiments have high computational demands and produce tremendous
amounts of data. In reaction to these demands new technologies such as grid and
semantic computing have been applied to astronomy leading to the fields of
astronomical grid computing and semantic astronomy. For example the grid
project of the German astronomy community (AstroGrid-D) (Enke et al. 2007), the
\textit{German Astrophysical Virtual Observatory} or the \textit{International
Virtual Observatory Alliance} have been active in these fields. Recent efforts
in semantic astronomy have focused on the usage of the Resource Description
Framework\footnote{http://www.w3.org/RDF/} (RDF) for metadata management. This
approach was also pursuit in AstroGrid-D with its information service
\textit{Stellaris} (H\"ogqvist, R\"oblitz \& Reinefeld 2007). The motivation is
to create a \textit{semantic web}, i.e. a collection of information with a
simple, more machine friendly structure. RDF is a new data model which
describes data through graphs of \textit{resources}, which are accessible
through Uniform Resource Identifiers (URIs). Complex resources are
hierarchically composed of simpler ones in analogy to the real-world object
they describe. For example a telescope is composed of a camera, which has a
filter wheel, which has filters, etc. This concept makes RDF an interesting
choice for the metadata management in heterogeneous software environments,
where automatic interaction between different components is desired. For
example a resource broker in a telescope network needs to find telescopes with
a certain filter.

However, a general problem arises when RDF is to be applied, since traditional
data formats are not RDF complaint. In astronomy for example information is
often stored in various XML dialects. Individual solutions for mapping a
particular dialect into RDF require a lot of effort and thus are often not
feasible. This constitutes a considerable barrier for the application of RDF to
this rich heritage of XML data. What is needed is a generic transformation to
map any XML dialect into RDF. This problem has been addressed before, e.g. in
the work by Trastour, Preist \& Coleman (2004), Battle (2006) and Akhtar et al.
(2008). However, their solutions are complex and require software which is not
yet available or does not work for the example discussed here. Another approach
is offered through the \textit{Virtuoso}
database\footnote{http://www.openlinksw.com/virtuoso/}. It provides means to
store XML data in relational databases and to dynamically convert it into RDF.
However, this conversion also requires a data specific schema and therefore can
not immediately be applied. Here a generic transformation for the automatic
conversion of XML into RDF is presented, which solves this problem for any XML
dialect.

To illustrate the problem and its solution the application to
OpenTel\footnote{http://www.gac-grid.org/project-products/RoboticTelescopes.html}
(Breitling, Granzer \& Enke 2008, Granzer et al. 2007), the grid integration
project for robotic telescopes of AstroGrid-D, is discussed. The goal of this
project is the integration of heterogeneous robotic telescopes into a common
infrastructure with standard components to build a global telescope network.
The infrastructure is provided by AstroGrid-D. The metadata management is
facilitated by the RDF information service Stellaris. Its purpose in a robotic
telescope network is to provide information e.g. about telescopes, weather and
scheduled observations. In OpenTel this information is exchanged in the Remote
Telescope Markup Language (RTML, Hessman 2006). RTML is an XML dialect
developed by the \textit{Heterogeneous Telescope Network} (Allan et al. 2006).
A transformation to RDF is needed to make the RTML data accessible to
Stellaris. Here a general transformation appears useful, since also other XML
data such as Usage
Records\footnote{http://staff.psc.edu/lfm/PSC/Grid/UR-WG/UR-Spec-gfd.58.pdf} of
grid jobs is provided to Stellaris e.g. for monitoring.

\section{Design}
A transformation to RDF has to create the URIs of its resources and connect
them through the RDF \textit{triple} structure consisting of subject, predicate
and object. However, this can be done in different ways which leaves room for
additional design criteria. The following have been added here:
\begin{itemize}
\item avoidance of blank nodes,
\item one-to-one mapping for a bidirectional extension,
\item independence of XML schema.
\end{itemize}
Blank nodes are subjects without name, i.e. URI. Therefore direct access to
them is not possible and some operations in RDF databases such as replacement
of nodes cannot be performed. Other disadvantages include reduced performance
of SPARQL processors and RDF databases. By giving every node a URI these
problems are avoided and direct access is possible.

One-to-one mapping means, that the transformation is injective and therefore
the original XML can uniquely be reconstructed from RDF through the inverse
transformation. A bidirectional transformation can be important e.g. in a
robotic telescope network where information about scheduled observations is
stored in RDF but where rescheduling requires the original RTML observation
request. Therefore in AstroGrid-D also the RTML observation requests were
stored along with the RDF. The inverse transformation could make this
additional service unnecessary. A unique reconstruction of the original XML
requires e.g. to preserve the distinction between attributes and elements. As
shown below, this is accomplished by the different transformation of attributes
and elements.

Independence from the XML schema means, that the transformation does not
require information contained in the XML schema. Therefore it can always
directly be applied and produce the same result. An example for information
only contained in the schema are XML sequences. XML sequences fix the order in
which elements have to appear. Therefore the order of elements might be
important for some parts of the data and thus should be preserved. It is then
also important for a later inverse transformation. A schema independent
transformation can only preserve or ignore the order information of all
elements. This transformation preserves this information by default, but it can
be disabled if it is not relevant for a particular data.

\section{Transformation}
In the following sections the transformation is explained for the simplified
RTML description of the robotic telescope STELLA-I (Strassmeier et al. 2004) in
listing \ref{lst:STELLA-I.rtml}.

RDF has different representations such as RDF/XML, N-Triples, N3 and Turtle.
For the following discussion the Turtle notation is used because of its compact
and clear structure. It represents triples in the order of subject, predicate
and object. Predicates and objects of the same subject can be grouped in a
block. Resources are enclosed by angle brackets, literals by quotation marks.
An example of the Turtle syntax is shown in listing \ref{lst:STELLA-I.rdf}. It
is the RDF counterpart to the RTML description generated by the presented
transformation.

Since the target is the transformation of XML dialects, XSLT
transformations\footnote{http://www.w3.org/TR/xslt/} are used. XSLT
transformations were specifically developed to facilitate the conversion from
one XML format to another. They are programmed through an XSLT stylesheet. The
stylesheet discussed here is name \texttt{xml2rdf3.xsl} and freely available
through the project
page\footnote{http://www.gac-grid.org/project-products/Software/XML2RDF.html}.
Its output is RDF/XML. The corresponding Turtle representation in listing
\ref{lst:STELLA-I.rdf} has been produced with the RDF parser \textit{Raptor}.

The transformation generates no base URI by default. Thus all resources have
relative URIs. Relative URIs are useful where a completion of the base URI is
done by an RDF service such as Stellaris. The RDF is then dereferencable for
RDF services of arbitrary URI. In contrast, RDF with absolute URIs is only
dereferencable for one particular service URI and might require a later
replacement. However, if desired a base URI can be added via the
\texttt{BaseURI} variable in the XSLT stylesheet.

\begin{figure*}[t]
  \centering \includegraphics[width=\textwidth]{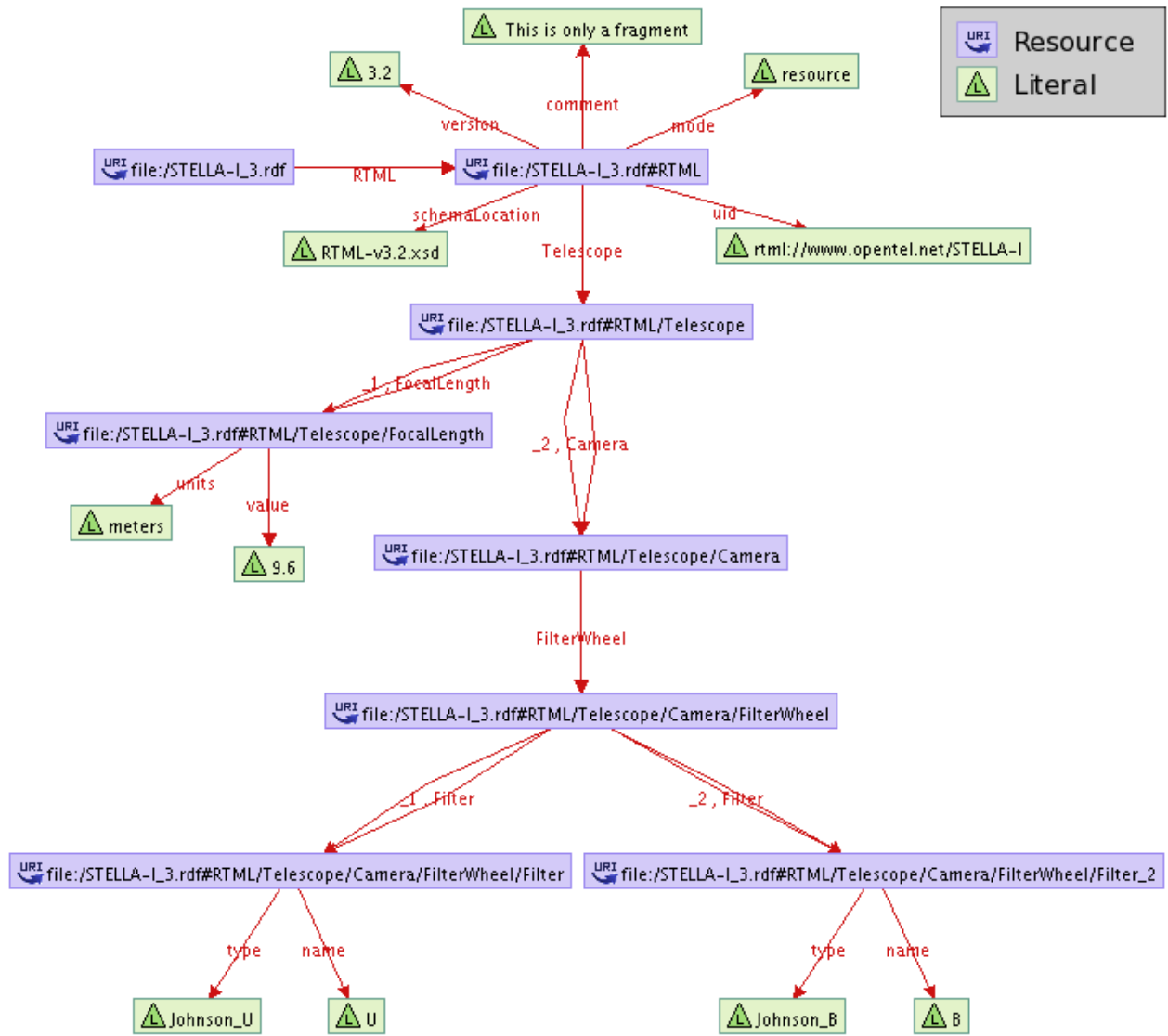}
  \caption{Graphical representation of the RDF graph of listing
    \ref{lst:STELLA-I.rdf} describing parts of the robotic telescope STELLA-I.
    Resources and literals are represented by purple and green labels,
    respectively (see legend top left). Labeled arrows represent predicates
    which connect subjects with objects to triples. The graphic was produced
    with \textit{RDF Gravity}. This RDF visualization tool also filled in the
    file name as base URI. The missing two backslashes after the colon is an
    unconventional feature of this tool.}
  \label{fig:RDFgraph}
\end{figure*}

The following sections explains the transformation of the different XML
components. A graphical representation is shown in Fig. \ref{fig:RDFgraph} to
support the discussion. It was obtained with the RDF graph visualization tool
\textit{RDF
Gravity}\footnote{http://semweb.salzburgresearch.at/apps/rdf-gravity/}.

\subsection{Transformation of Attributes}
XML attributes have a natural correspondence to triples since they contain two
pieces of information. This suggests their transformation into predicate and
object of the parent element which itself serves as subject. For example the
\texttt{units} attribute of the \texttt{FocalLength} element at line 7 of
listing \ref{lst:STELLA-I.rtml} transforms into the triple at line 23-24 of
listing \ref{lst:STELLA-I.rdf}.

\subsection{Transformation of Text}
A natural representation of XML text is provided by RDF literals. Literals can
be further specified by type and language information, but this is not used
here. Literals can only serves as objects. The subject is already provided by
the parent element as defined by the attribute transformation above. The
predicate can be provided by the RDF utility class \texttt{rdf:value}.
Again the \texttt{FocalLength} element at line 7 of listing
\ref{lst:STELLA-I.rtml} serves as an example. Its text \texttt{9.6} transforms
into the triple at line 23 and 25 of listing \ref{lst:STELLA-I.rdf}.

\subsection{Transformation of Elements}
Child elements cannot be modeled with literal objects as before but require
resources. The necessary URIs are constructed from the sequence of ancestor
elements. In case of child elements with the same name, an ID is also appended,
to make the URIs unique. The parent element provides the predicate. The
\texttt{FilterWheel} and its two \texttt{Filter} elements line 9-12 of listing
\ref{lst:STELLA-I.rtml} serve as example. The filter elements result in two
resource at line 35-41 of listing \ref{lst:STELLA-I.rdf}. A \texttt{2} is
appended as unique ID to the second URI separated by an underscore.

In addition to the predicate from the parent element (line 30-31), another
predicate (\texttt{rdf:\_no}) is added to conserve the element's order (line
32-33). However this is only added where necessary, i.e. to elements with
siblings. It can be disabled (commented out) in the XSLT stylesheet if it is
not relevant for the XML data.

\subsection{Transformation of Comments}
XML comments are preserved in additional triples to the element where they
occur. \texttt{xs:comment} serves as predicate. An example is given at line 5
of listing \ref{lst:STELLA-I.rtml} and the corresponding lines 6 and 12 of
listing \ref{lst:STELLA-I.rdf}.

\begin{lstlisting}[float=t, caption={Description of robotic telescope STELLA-I.},
    label={lst:STELLA-I.rtml}]
<?xml version="1.0" encoding="UTF-8"?>
<RTML version="3.2" mode="resource" uid="rtml://www.opentel.net/STELLA-I"
  xmlns:xsi="http://www.w3.org/2001/XMLSchema-instance"
  xmlns="http://www.rtml.org/v3.2" xsi:schemaLocation="RTML-v3.2.xsd">
  <!-- This is only a fragment -->
  <Telescope>
    <FocalLength units="meters">9.6</FocalLength>
    <Camera>
      <FilterWheel>
        <Filter type="Johnson_U" name="U"/>
        <Filter type="Johnson_B" name="B"/>
      </FilterWheel>
    </Camera>
  </Telescope>
</RTML>
\end{lstlisting}

\subsection{Execution}
The application of the XSLT stylesheet requires an XSLT version 1.0 compliant
processor such as \textit{xsltproc}. The command line syntax is:

\begin{lstlisting}[numbers=none, frame=]
xsltproc xml2rdf3.xsl STELLA-I.rtml
\end{lstlisting}

\section{Conclusion}
A generic transformation for XML data into RDF/XML has been presented. Owing to
its generality it can be applied to 
any XML dialect and serve as standard to automate the
\begin{lstlisting}[caption={RDF (Turtle) representation
    of listing \ref{lst:STELLA-I.rtml}.},
    label={lst:STELLA-I.rdf}]
@prefix rdf: <http://www.w3.org/1999/02/22-rdf-syntax-ns#> .
@prefix ns1: <http://www.rtml.org/v3.2#> .
@prefix xsi: <http://www.w3.org/2001/XMLSchema-instance#> .
@prefix xs: <http://www.w3.org/TR/2008/REC-xml-20081126#> .

<file:///STELLA-I_3.rdf#RTML>
    ns1:Telescope <file:///STELLA-I_3.rdf#RTML/Telescope> ;
    ns1:mode "resource" ;
    ns1:uid "rtml://www.opentel.net/STELLA-I" ;
    ns1:version "3.2" ;
    xsi:schemaLocation "RTML-v3.2.xsd" ;
    xs:comment " This is only a fragment " .

<file:///STELLA-I_3.rdf>
    ns1:RTML <file:///STELLA-I_3.rdf#RTML> .

<file:///STELLA-I_3.rdf#RTML/Telescope>
    ns1:Camera <file:///STELLA-I_3.rdf#RTML/Telescope/Camera> ;
    ns1:FocalLength <file:///STELLA-I_3.rdf#RTML/Telescope/FocalLength> ;
    rdf:_1 <file:///STELLA-I_3.rdf#RTML/Telescope/FocalLength> ;
    rdf:_2 <file:///STELLA-I_3.rdf#RTML/Telescope/Camera> .

<file:///STELLA-I_3.rdf#RTML/Telescope/FocalLength>
    ns1:units "meters" ;
    rdf:value "9.6" .

<file:///STELLA-I_3.rdf#RTML/Telescope/Camera>
    ns1:FilterWheel <file:///STELLA-I_3.rdf#RTML/Telescope/Camera/FilterWheel> .

<file:///STELLA-I_3.rdf#RTML/Telescope/Camera/FilterWheel>
    ns1:Filter <file:///STELLA-I_3.rdf#RTML/Telescope/Camera/FilterWheel/Filter>, <file:///STELLA-I_3.rdf#RTML/Telescope/Camera/FilterWheel/Filter_2> ;
    rdf:_1 <file:///STELLA-I_3.rdf#RTML/Telescope/Camera/FilterWheel/Filter> ;
    rdf:_2 <file:///STELLA-I_3.rdf#RTML/Telescope/Camera/FilterWheel/Filter_2> .

<file:///STELLA-I_3.rdf#RTML/Telescope/Camera/FilterWheel/Filter>
    ns1:name "U" ;
    ns1:type "Johnson_U" .

<file:///STELLA-I_3.rdf#RTML/Telescope/Camera/FilterWheel/Filter_2>
    ns1:name "B" ;
    ns1:type "Johnson_B" .
\end{lstlisting}
transformation of XML data into RDF. Thus it solves this
conversion problem of semantic computing. Its application is simple and has
proven useful to semantic astronomy and grid computing. The transformation is
injective so that the inverse transformation as bidirectional extension can be
developed.

\acknowledgement I would like to thank my colleagues Mikael H\"ogqvist, Harry
Enke and Stevan White for helpful discussion and useful comments. Moreover many
thanks goes to Harry Enke, Matthias Steinmetz and the German \textit{Federal
  Ministry of Education and Research} (BMBF) for supporting this work within
AstroGrid-D and the D-Grid initiative.

\addcontentsline{toc}{section}{References}
\small

\onecolumn 
\begin{lstlisting}[caption={XSLT stylesheet for the
 transformation from XML to RDF as found at the project page at
 http://www.gac-grid.org/project-products/Software/XML2RDF.html. It is not part
 of the publication in Astronomical Notes. Only the reference to the project
 page is included.}, label={lst:xml2rdf3.xsl}]
<?xml version="1.0" encoding="UTF-8"?>
<!-- xml2rdf3.xsl        XSLT stylesheet to transform XML into RDF/XML

     Version             3.0  (2009-05-28)
     Changes to V2.5     rdf:value for all text, no attribute triples,
                         order predicates, comments as triples
     Web page            http://www.gac-grid.org/project-products/Software/XML2RDF.html
     Usage               xsltproc xml2rdf3.xsl file.xml
     Author              Frank Breitling (fbreitling at aip.de)
     Copyright 2009      AstroGrid-D (http://www.gac-grid.org/) 
   
     Licensed under the Apache License, Version 2.0 (the "License");
     you may not use this file except in compliance with the License.
     You may obtain a copy of the License at
 
         http://www.apache.org/licenses/LICENSE-2.0
 
     Unless required by applicable law or agreed to in writing, software
     distributed under the License is distributed on an "AS IS" BASIS,
     WITHOUT WARRANTIES OR CONDITIONS OF ANY KIND, either express or implied.
     See the License for the specific language governing permissions and
     limitations under the License. -->

<xsl:stylesheet version="1.0" xmlns:xsl="http://www.w3.org/1999/XSL/Transform"
       xmlns:rdf="http://www.w3.org/1999/02/22-rdf-syntax-ns#"
       xmlns:xs="http://www.w3.org/TR/2008/REC-xml-20081126#">

       <xsl:strip-space elements="*"/>
       <xsl:output method="xml" indent="yes"/>

       <xsl:param name="BaseURI"><!-- define if not completed by application-->
              <!-- e.g. http://www.exampleURI.net/STELLA-I -->
       </xsl:param>

       <!-- Begin RDF document -->
       <xsl:template match="/">
              <xsl:element name="rdf:RDF">
                     <rdf:Description>
                            <xsl:attribute name="rdf:about"/>
                            <xsl:apply-templates select="/*|/@*"/>
                     </rdf:Description>
              </xsl:element>
       </xsl:template>

       <!-- Create triples for elements, turn elements into resources -->
       <xsl:template match="*">
              <xsl:param name="subjectname"/>

              <!-- Build URI from ancestor elements -->
              <xsl:variable name="newsubjectname">
                     <xsl:if test="$subjectname=''">
                            <xsl:value-of select="$BaseURI"/>
                            <xsl:text>#</xsl:text>
                     </xsl:if>
                     <xsl:value-of select="$subjectname"/>
                     <xsl:value-of select="name()"/>
                     <!-- Add unique ID for sibling elements with same tag -->
                     <xsl:variable name="number">
                            <xsl:number/>
                     </xsl:variable>
                     <xsl:if test="$number > 1">
                            <xsl:text>_</xsl:text>
                            <xsl:number/>
                     </xsl:if>
              </xsl:variable>

              <xsl:element name="{name()}" namespace="{concat(namespace-uri(),'#')}">
                     <rdf:Description>
                            <xsl:attribute name="rdf:about">
                                   <xsl:value-of select="$newsubjectname"/>
                            </xsl:attribute>
                            <xsl:apply-templates select="@*|node()">
                                   <xsl:with-param name="subjectname"
                                          select="concat($newsubjectname,'/')"/>
                            </xsl:apply-templates>
                     </rdf:Description>
              </xsl:element>

              <!-- Create ordering triples with rdf:_no predicates, 
                   comment out if not needed -->
              <xsl:if test="count(../*) >1">
                     <xsl:element name="{concat('rdf:_',count(preceding-sibling::*)+1)}">
                            <rdf:Description>
                                   <xsl:attribute name="rdf:about">
                                          <xsl:value-of select="$newsubjectname"/>
                                   </xsl:attribute>
                            </rdf:Description>
                     </xsl:element>
              </xsl:if>
       </xsl:template>

       <!-- Create triples for attributes -->
       <xsl:template match="@*" name="attributes">
              <xsl:variable name="ns">
                     <!-- If attribute doesn't have a namespace use element namespace -->
                     <xsl:choose>
                            <xsl:when test="namespace-uri()=''">
                                   <xsl:value-of select="concat(namespace-uri(..),'#')"/>
                            </xsl:when>
                            <xsl:otherwise>
                                   <xsl:value-of select="concat(namespace-uri(),'#')"/>
                            </xsl:otherwise>
                     </xsl:choose>
              </xsl:variable>
              <xsl:element name="{name()}" namespace="{$ns}">
                     <xsl:value-of select="."/>
              </xsl:element>
       </xsl:template>

       <!-- Create triples for text with rdf:value predicates -->
       <xsl:template match="text()">
              <xsl:element name="rdf:value">
                     <xsl:value-of select="."/>
              </xsl:element>
       </xsl:template>

       <!-- Create triples for comments -->
       <xsl:template match="comment()">
              <xsl:element name="xs:comment">
                     <xsl:value-of select="."/>
              </xsl:element>
       </xsl:template>

</xsl:stylesheet>
\end{lstlisting}

\end{document}